\documentstyle{article}
\newcommand{\BA}{$B\to Al\nu_l$}
\newcommand{\BAV}{$B\to A^*l\nu_l$}
\newcommand{\qm}{q^2_{max}}

\begin{document}
\begin{center}
\Large \bf Relativistic Description of Exclusive Semileptonic Decays
of Heavy Mesons
\end{center}
\vspace{1.2cm}
\begin{center}
R.N.Faustov, V.O.Galkin and A.Yu.Mishurov
\end{center}
\begin{center}
\small
\it
Russian Academy of Sciences, Scientific Council for Cybernetics,
Vavilov Street
40, Moscow, 117333, Russia.
\end{center}
\vspace{1.2cm}
\begin{abstract}
Using quasipotential approach, we have studied exclusive semileptonic
decays of heavy mesons with the account of relativistic effects. Due
to more
complete relativistic description of the $s$ quark more precise
expressions for semileptonic form factors are obtained. Various
differential distributions in exclusive semileptonic decays of heavy
mesons are calculated.  It is argued that consistent account of
relativistic effects and HQET motivated choice of the parameters of
quark-antiquark potential allow to get reliable value for the ratio
$A_2(0)/A_1(0)$ in the $D\to K^*l\nu_l$ decay as well as the
ratio~$\Gamma(D\to K^*l\nu_l)/\Gamma(D\to Kl\nu_l)$. All calculated
branching
ratios are in accord with available experimental data.  \\
\noindent PACS number(s): 12.39Ki, 13.20He, 13.40Hq
\end{abstract}

\section{Introduction} Semileptonic decays of heavy mesons provide an
important tool to investigate quark dynamics and to determine
Cabibbo-Kobayshi-Maskawa (CKM) matrix elements.  Hadron dynamics is
contained in form factors, which are Lorentz invariant functions of
$q^2$,
the square of momentum transfer. These form factors cannot be
calculated
from the first principles of QCD by now. Thus various potential
models, sum
rules and lattice calculations have been proposed~\cite{1}--\cite{6}.
 Recently
considerable progress has been achieved in describing heavy meson
decays by
the use of heavy quark effective theory (HQET)~\cite{7}. It has been
found
that in the limit of infinitely heavy $b$ and $c$ quarks their mass
and spin decouple
from the dynamics of the decay and the description of a process such
as
$B\to Dl\nu_l$ is strongly simplified. For $D$ decays HQET
predictions are less
useful, because in this case symmetry breaking corrections appear to
be
rather large.  It is also important to note that since $B$ and $D$
mesons
contain light quark, relativistic effects are quite significant and
consistent relativistic description of heavy-light quark system is
necessary.

Our relativistic quark model (RQM), has some features that make it
attractive and reliable for the description of heavy mesons. Firstly,
RQM provides a consistent scheme for calculation of all relativistic
corrections and allows for the heavy quark $1/m_Q$ expansion.
Secondly, it
has been found~\cite{8} that the general structure of leading, next-
to-leading
and second order $1/m_Q$ corrections in RQM is in accord with the
predictions of HQET. The heavy quark symmetry and QCD impose some
rigid
constraints on the parameters of the long-range confining potential
of our
model. It gives  an additional motivation for the choice of the main
parameters of RQM and leads us to the conclusion that the confining
quark-antiquark potential in meson is predominantly Lorentz-vector
(with
the Pauli term), while the scalar potential is anticonfining and
helps to
reproduce the initial nonrelativistic potential. This model has been
applied to the calculations of meson mass spectra~\cite{9}, radiative
decay
widths~\cite{10}, pseudoscalar decay constants~\cite{11}, rare
radiative~\cite{12} and nonleptonic~\cite{13} decay rates.
Semileptonic decays
of $B$ and $D$ mesons have been considered in our model in~\cite{14}.
 Here
we refine our previous analysis with more complete account of
relativistic
effects and HQET constraints. We also consider exclusive decay
spectra and
$q^2$ dependence of form factors.

In Sect.2 we briefly describe RQM. Sect.3 is devoted to the
calculation of
form factors and semileptonic branching ratios and contains analytical
expressions and numerical results for the differential distributions
for
the decays into pseudoscalar as well as vector final states. We give
our
conclusions in Sect.4.

\section{Relativistic Quark Model} Our model is based on the
quasipotential
approach in quantum field theory~\cite{15}.  A quark-antiquark bound
system
with the mass $M$ and relativistic momentum {\bf p} in the center of
mass
system
is described by  a single-time quasipotential wave function
$\Psi_M({\bf
p})$, projected onto positive-energy states. This wave function
satisfies
the quasipotential equation
\begin{equation} \label{1} \left[M^2-({\bf
p}^2+m_1^2)^{1/2}-({\bf p}^2+m_2^2)^{1/2}\right] \Psi_M({\bf p}) =
\int
\frac{d^3{\bf q}}{(2\pi)^3} V({\bf p},{\bf q};M)\Psi_M({\bf q}),
\end{equation}
The quasipotential equation (\ref{1}) can be transformed
into a local Schr\"odinger-like equation~\cite{16}
\begin{equation}
\label{2} \left[\frac{b^2(M)}{2\mu_R}-\frac{{\bf p}^2}{2\mu_R}\right]
\Psi_M({\bf p}) = \int \frac{d^3{\bf q}}{(2\pi)^3} V({\bf p},{\bf
q};M)\Psi_M({\bf q}),
\end{equation}
where the relativistic reduced mass
is
\begin{equation} \label{3}
\mu_R=\frac{E_1E_2}{E_1+E_2}=\frac{M^4-(m_1^2-m_2^2)^2}{4M^3};
\end{equation}
$$ E_1=\frac{M^2-m_2^2+m_1^2}{2M};\qquad
E_2=\frac{M^2-m_1^2+m_2^2}{2M};\qquad E_1+E_2=M.  $$
and the square of the
relative momentum on the mass shell is
\begin{equation}
\label{4}
b^2(M)=
\frac{\left[M^2-(m_1+m_2)^2\right]\left[M^2-(m_1-m_2)^2\right]}{4M^2},
\end{equation}
$m_{1,2}$ are the quark masses.

Now it is necessary to
construct the quasipotential $V({\bf p}, {\bf q}; M)$ of the
quark-antiquark interaction. As well known from QCD,
in view of the property of asymptotic freedom the one-gluon exchange
potential gives the main contribution at short distances. With the
increase of the distance the
long-range confining interaction becomes dominant. At present the
form of
this interaction cannot be established in the framework of QCD. The
most
general kernel of $q\bar q$ interaction, corresponding to the
requirements
of Lorentz invariance and of $P$ and $T$ invariance,
contains~\cite{17}, \cite{18}
scalar, pseudoscalar, vector, axial-vector and tensor parts. The
analysis
carried out in~\cite{9}, \cite{17} has shown that the leading
contributions to the
confining part of the potential should have a vector and scalar
structure.
On the basis of these arguments we have assumed that the effective
interaction is the sum of the one-gluon exchange term and the mixture
of
long-range vector with scalar potentials. We have also assumed that
at large
distances quarks acquire universal nonperturbative anomalous
chromomagnetic
moments and thus the vector long-range potential contains the Pauli
interaction. The quasipotential is defined by~\cite{9}
\begin{equation}
\label{5}
V({\bf p}, {\bf q}; M)=\bar u_1(p)\bar u_2(-p)\left(\frac{4}{3}\alpha
_sD_{\mu\nu}({\bf k})\gamma_1^{\mu}\gamma_2^{\nu}+V_{conf}^V({\bf
k})\Gamma_1
^{\mu}\Gamma_{2;\mu}+ V_{conf}^S({\bf k})\right)u_1(q)u_2(-q),
\end{equation}
where $\alpha_s$ is the QCD coupling constant, $D_{\mu\nu}$ is the
gluon
propagator; $\gamma_\mu$ and $u(p)$ are the Dirac matrices and
spinors;
${\bf k}={\bf p}-{\bf q};$ $\Gamma_\mu$ is the effective vector
vertex at large
distances,
\begin{equation}
\label{6}
\Gamma_\mu({\bf k})=\gamma_\mu+\frac{i\kappa}{2m}\sigma_{\mu\nu}k^\nu,
\end{equation}
$\kappa$ is the anomalous chromomagnetic quark moment.

The complete expression for the quasipotential obtained from
(\ref{5}), (\ref{6})
with the account of the relativistic corrections of order $v^2/c^2$
can be found in~\cite{9}.
In the nonrelativistic limit vector and scalar
confining potentials reduce to
\begin{equation}
\label{7}
V_{conf}^V(r)=(1-\varepsilon)(Ar+B), \qquad
V_{conf}^S(r)=\varepsilon(Ar+B),
\end{equation}
reproducing $V_{nonrel}^{conf}(r)=V_{conf}^S+V_{conf}^V=Ar+B$, where
$\varepsilon$
is the mixing coefficient.

All the parameters of our model: quark masses, parameters of linear
confining
potential $A$ and $B$, mixing coefficient $\varepsilon$ and anomalous
chromomagnetic quark moment $\kappa$ were originally fixed from the
analysis
of meson masses~\cite{9} and radiative decays~\cite{10}. Quark
masses: $m_b$=~4.88~
GeV; $m_c$=~1.55~GeV; $m_s$=~0.50~GeV; $m_{u,d}$=~0.33~GeV and
parameters of
the linear potential: $A$=~$0.18$~GeV$^2$; B=~$-0.30$~GeV have
standard values for quark
models. The value of mixing coefficient of vector and scalar confining
potentials $\varepsilon= -0.9$ has been primarily chosen from the
consideration
of meson radiative decays, which are rather sensitive to the Lorentz-
structure
of the confining potential~\cite{10}. Universal anomalous
chromomagnetic moment
of quarks $\kappa= -1$ has been fixed from the analysis of the fine
splitting
of heavy quarkonia $^3P_J$ states~\cite{9}.

Recently, in the framework of RQM, the $1/m_Q$ expansion of the
matrix elements of
weak currents between pseudoscalar and vector heavy meson states
has been performed~\cite{8}. It has been found
that the particular structure of $1/m_Q$ corrections up to the second
order
predicted by HQET can be reproduced in RQM only with some specific
values of
$\kappa$ and $\varepsilon$. The analysis of the first order
corrections~\cite{8}
allowed to fix $\kappa=-1$, while from the consideration of the
second order
corrections it has been obtained that mixing parameter $\varepsilon$
should be
$\varepsilon=-1$. Thus HQET, and hence QCD, imposes strong
constraints on the
parameters of the long-range confining potential. The obtained value
of
$\varepsilon$ is very close to the previous one, determined
phenomenologically
from radiative decays~\cite{10} and the value of $\kappa$ coincides
with
the result, obtained from the mass spectra~\cite{9}. Therefore, there
is an
important QCD and heavy quark symmetry motivation for the choice of
the main
parameters of our model: $\varepsilon=-1$, $\kappa=-1$.

\section{Exclusive Semileptonic Decay}
\subsection{Form Factors and Decay Widths}
For semileptonic decay $B\to A(A^*)l\nu_l$ of pseudoscalar meson $B$
into
pseudoscalar (vector) meson $A(A^*)$  the
differential width can be written as
\begin{equation}
\label{8}
d\Gamma(B\to A(A^*)l\nu_l)=\frac{1}{2M_B}\mid {\cal A}(B\to
A(A^*)l\nu_l)\mid ^2d\Phi,
\end{equation}
where
\begin{equation}
\label{9}
d\Phi=(2\pi)^4\delta^{(4)}\left(p_B-p_l-p_{\nu_l}-
p_A\right)\prod_i\frac{d^3p_i}
{(2\pi)^32E_i},\qquad i=A,\,l,\,\nu_l,
\end{equation}
$p_B$ is four-momentum of initial meson, $p_A$ is four-momentum of
final meson,
$p_l$ and $p_{\nu_l}$ are four-momenta of lepton and neutrino
respectively.

The relevant transition amplitude looks like
\begin{equation}
\label{10}
{\cal A}(B\to A(A^*)l\nu_l)=\langle
A(A^*)l\nu_l|H_{eff}|B\rangle=\frac{G_F}{\sqrt{2}}
V_{ab}L_{\mu}H^{\mu},
\end{equation}
where
\begin{equation}
\label{11}
H_{eff}=\frac{G_F}{\sqrt{2}}J_{hadron}^{\mu}J_{lepton;\mu},
\end{equation}
$V_{ab}$ is the CKM matrix element
connected with $b\to a$~transition.

The leptonic $L_\mu$ and hadronic $H_\mu$ currents are defined by
\begin{equation}
\label{12}
L_{\mu}=\bar l\gamma_\mu(1-\gamma_5)\nu_l,
\end{equation}
\begin{equation}
\label{13}
H_{\mu}=\langle A(A^*)|\bar a\gamma_\mu(1-\gamma_5)b|B\rangle,
\end{equation}
the initial meson $B$ has the quark structure $(b\bar q)$ and the
final
meson $A(A^*)$ has the quark structure~$(a\bar q)$.

The matrix element of hadron current can be expressed in terms of
Lorentz-invariant
form factors

a) For $0^-\to 0^-$ transition \BA
\begin{equation}
\label{14}
\langle A(p_A)|J^V_\mu|B(p_B)\rangle=f_+(q^2)(p_A+p_B)_{\mu}+f_-
(q^2)(p_B-p_A)_{\mu};
\end{equation}

b) For $0^-\to 1^-$ transition \BAV
\begin{eqnarray}
\label{15}
\langle
A^*(p_A,e)|J^V_\mu|B(p_B)\rangle&=&i\frac{V(q^2)}{M_A+M_B}\epsilon
_{\mu\nu\rho\sigma}e^{*\nu}(p_A+p_B)^{\rho}(p_B-p_A)^{\sigma};\\
\label{16}
\nonumber
\langle A^*(p_A,e)|J^A_\mu|B(p_B)\rangle&=&(M_A+M_B)A_1(q^2)e^*_\mu
-\frac{A_2(q^2)}{M_A+M_B}(e^*p_B)(p_A+p_B)_{\mu}\\
&&+\frac{A_3(q^2)}{M_A+M_B}(e^*p_B)(p_B-p_A)_{\mu};
\end{eqnarray}
where $q=p_B-p_A$; $J_\mu^V=(\bar a\gamma_\mu b)$ and
$J_\mu^A=(\bar a\gamma_\mu\gamma_5 b)$ are vector and axial parts of
the
weak current; $e_\mu$ is the polarization vector of $A^*$ meson.

Since $q=p_l+p_{\nu_l}$, the terms proportional to $q_\nu$, i.e. $f_-
$ and
$A_3$ give contributions proportional to the lepton masses and do not
influence
significantly the transition amplitude, except for the case of heavy
$\tau$
lepton, and thus will not be considered.

The matrix element of the local current $J$ between bound states in
the
quasipotential method has the form~\cite{19}
\begin{equation}
\label{25}
\langle A|J_\mu(0)|B\rangle=\int\frac{d^3{\bf p}d^3{\bf q}}{(2\pi)^6}
\bar\Psi_A({\bf p})\Gamma_\mu({\bf p}, {\bf q})\Psi_B({\bf q}),
\end{equation}
where $\Gamma_\mu({\bf p}, {\bf q})$ is the two-particle vertex
function and
$\Psi_{A,B}$ are meson wave functions projected onto the positive
energy
quark states.

In the case of semileptonic decays $J_\mu=J_{hadron;\mu}=J_\mu^V-
J_\mu^A$
is the weak quark current and in order to calculate its matrix element
between meson states it is necessary to consider the contributions to
$\Gamma$
from Figs.~1 and 2. The vertex functions obtained from these diagrams
look
like
\begin{equation}
\label{26}
\Gamma_\mu^{(1)}({\bf p}, {\bf q})=\bar u_a(p_1)\gamma_\mu(1-
\gamma_5)u_b(q_1)
(2\pi)^3\delta({\bf p}_2-{\bf q}_2),
\end{equation}
and
\begin{eqnarray}
\label{27}
\nonumber
\Gamma_\mu^{(2)}({\bf p}, {\bf q})&=&\bar u_a(p_1)\bar
u_q(p_2)\Big[\gamma_{1\mu}
(1-\gamma_1^5)\frac{\Lambda_b^{(-)}({\bf
k})}{\varepsilon_b(k)+\varepsilon_b(p_1)}
\gamma_1^0V({\bf p}_2-{\bf q}_2)\\
&&+V({\bf p}_2-{\bf q}_2)\frac{\Lambda_a^{(-)}({\bf
k}')}{\varepsilon_a(k')+\varepsilon_a(q_1)}
\gamma_1^0\gamma_{1\mu}(1-\gamma_1^5)\Big] u_b(q_1)u_q(q_2),
\end{eqnarray}
where ${\bf k}={\bf p}_1-{\bf \Delta}$;\qquad ${\bf k'}={\bf
q}_1+{\bf \Delta}$;
\qquad ${\bf \Delta}={\bf p}_B-{\bf p}_A$;\qquad
$\varepsilon(p)=(m^2+{\bf p}^2)^{1/2}$;
$$
\Lambda^{(-)}(p)=\frac{\varepsilon(p)-(m\gamma^0+\gamma^0{\vec
\gamma}{\bf p})}
{2\varepsilon(p)}.
$$
As one can see, the form of relativistic corrections resulting from
$\Gamma_\mu^{(2)}({\bf p}, {\bf q})$
is explicitly dependent on the Lorentz-structure of quark-antiquark
potential.

Our previous analysis of the semileptonic $B\to D(D^*)$ and $D\to
K(K^*)$
transitions~\cite{14} was based on the assumption that we could
expand (\ref{27})
up to the order ${\bf p}^2/m^2$ with respect to both $b$ and $a$
quarks. This assumption proved to be quite adequate in the case of
$B\to Dl\nu_l$  where both
$b$ and $c$ quarks are heavy. However, the final $s$ quark  is not
heavy enough.
It would  be more
accurate not to expand $\Gamma_\mu^{(2)}({\bf p}, {\bf q})$ at all,
but one should
do it in order to perform one of the integrations in (\ref{25}). Our
statement is
that more reliable results for semileptonic $D\to K$ decays can be
obtained
by using
${\bf p}^2/\varepsilon_a^2(p)$ expansion instead of ${\bf p}^2/m^2$
in (\ref{27}).

It is also necessary to take into account that the wave function of
final $A$ meson
$\Psi_{A,{\bf p}_A}({\bf p})$ is connected with one in $A$ rest frame
$\Psi_{A,{\bf 0}}({\bf p})$ as follows~\cite{19}
\begin{equation}
\label{28}
\Psi_{A,{\bf p}_A}({\bf p})=D^{1/2}_a(R^W_{L{\bf
p}_A})D^{1/2}_q(R^W_{L{\bf p}_A})
\Psi_{A,{\bf 0}}({\bf p}),
\end{equation}
where $D^{1/2}(R)$ is well-known rotation matrix and $R^W$ is the
Wigner
rotation.

The meson functions in the rest frame have been calculated~\cite{20}
by numerical
solution of the quasipotential equation~(\ref{2}). However, it is
more convinient
to use analytical expressions for meson wave functions. The
examination of
numerical results for the ground state wave functions of mesons
containing
at least one light quark has shown that they can be well approximated
by the Gaussian functions
\begin{equation}
\label{29}
\Psi_M({\bf p})\equiv\Psi_{M,{\bf 0}}({\bf
p})=\left(\frac{4\pi}{\beta_M^2}\right)^{3/4}
\exp\left(-\frac{{\bf p}^2}{2\beta_M^2}\right),
\end{equation}
with the deviation less than 5\%.

The parameters are
\begin{eqnarray}
\label{30}
\nonumber
&&\beta_B=0.41 \,\mbox {GeV};\qquad \beta_K=\beta_{K^*}=0.33 \,\mbox
{GeV};
\qquad \beta_{\phi}=0.36 \,\mbox{GeV}; \\
&&\beta_D=0.38 \,\mbox {GeV};\qquad \beta_{D_s}=0.44 \,\mbox{GeV}
\end{eqnarray}

In the $B$ meson rest frame equations~(\ref{14})--(\ref{16}) can be
written
in the three-dimensional form

a) For $0^-\to 0^-$ transition \BA
\begin{eqnarray}
\label{17}
\langle A(p_A)|J^V_0|B(p_B)\rangle&=&f_+(q^2)(M_B+E_A)+f_-(q^2)(M_B-
E_A),\\
\label{18}
\langle A(p_A)|{\bf J}^V|B(p_B)\rangle&=&(f_-(q^2)-f_+(q^2)){\bf
\Delta},
\end{eqnarray}

b) For $0^-\to 1^-$ transition \BAV
\begin{eqnarray}
\label{19}
\langle
A^*(p_A,e)|J^V_0|B(p_B)\rangle&=&i\frac{V(q^2)}{M_A+M_B}\epsilon
_{0\nu\rho\sigma}\tilde e^{*\nu}(p_A+p_B)^{\rho}(p_B-
p_A)^{\sigma}=0,\\
\label{20}
\langle A^*(p_A,e)|{\bf
J}^V|B(p_B)\rangle&=&i\frac{2M_B}{M_A+M_B}V(q^2)\left[
\tilde {\bf e}^*{\bf \Delta}\right],\\
\label{21}
\nonumber
\langle A^*(p_A,e)|J^A_0|B(p_B)\rangle&=&{\tilde
e}_0^*\Bigl(A_1(q^2)(M_A+M_B)
-A_2(q^2)\frac{M_B}{M_A+M_B}(M_B+E_A)\\
&&+A_3(q^2)\frac{M_B}{M_A+M_B}(M_B-E_A)\Bigr),\\
\label{22}
\langle A^*(p_A,e)|{\bf J}^A|B(p_B)\rangle&=&A_1(q^2)(M_A+M_B)\tilde
{\bf e}-
{\bf \Delta}\tilde e_0^*\frac{M_B}{M_A+M_B}(A_2(q^2)+A_3(q^2)),
\end{eqnarray}
where $e_\mu=(0,{\bf e)}$ is the polarization vector
of $A^*$ meson in its rest frame, $\tilde e_\mu$ is the vector
obtained from
$e_\mu$ by the Lorentz transformation $L_{\bf \Delta}$
\begin{equation}
\label{23}
\tilde e_\mu=L_{\bf \Delta}e_\mu.
\end{equation}

The components of $\tilde e_\mu$ look like
\begin{equation}
\label{24}
\tilde e_0=\frac{{\bf e \Delta}}{M_A}, \qquad
\tilde {\bf e}={\bf e}+\frac{{\bf \Delta}({\bf \Delta
e})}{M_A(E_A+M_A)}=
    {\bf e}+\tilde e_0\frac{{\bf \Delta}}{E_A+M_A}.
\end{equation}
Equations (\ref{17}), (\ref{18}) and (\ref{19})--(\ref{22})
determine form factors $f_+$, $f_-$ and $V$, $A_1$, $A_2$, $A_3$
respectively.

Substituting the vertex functions~(\ref{26}) and~(\ref{27}), with the
account
of wave function transformation~(\ref{28}) and quasipotential
equation~(\ref{1})
in the matrix element~(\ref{25}) and using eqs.~(\ref{17}),
(\ref{18}) and (\ref{19})--(\ref{22}) we get the following expressions
at $q^2=q^2_{max}=(M_B-M_A)^2$ point
\begin{eqnarray}
\label{31}
f_+(\qm)&=&f_+^{(1)}(\qm)+\varepsilon f_{+S}^{(2)}(\qm)+(1-
\varepsilon)
f_{+V}^{(2)}(\qm),\\
\label{32}
A_1(\qm)&=&A_1^{(1)}(\qm)+\varepsilon A_{1S}^{(2)}(\qm)+(1-
\varepsilon)
A_{1V}^{(2)}(\qm),\\
\label{33}
A_2(\qm)&=&A_2^{(1)}(\qm)+\varepsilon A_{2S}^{(2)}(\qm)+(1-
\varepsilon)
A_{2V}^{(2)}(\qm),\\
\label{34}
V(\qm)&=&V^{(1)}(\qm)+\varepsilon V_{S}^{(2)}(\qm)+(1-\varepsilon)
V_{V}^{(2)}(\qm),
\end{eqnarray}
where $f_+^{(1)}$, $f_{+S,V}^{(2)}$, $A_{1,2}^{(1)}$, $A_{1S,V}^{2}$,
$A_{2S,V}^{(2)}$, $V^{(1)}$ and $V_{S,V}^{(2)}$ are given in Appendix
A.
In (\ref{31})--(\ref{34}) indeces $(1)$ and $(2)$ correspond to the
diagrams
in Figs.~1 and 2, $S$ and $V$ correspond to the scalar and vector
potentials
of quark-interaction.

Now our concern is to find $q^2$ dependence of the form factors.
Components of
axial and vector currents can be expressed in terms of two functions
$F_1({\bf \Delta})$ and $F_2({\bf \Delta})$, ${\bf \Delta}={\bf p}_B-
{\bf p}_A$,
\begin{eqnarray}
\label{46}
J_0^V({\bf \Delta})&=&F_2({\bf \Delta}),\\
\label{47}
{\bf J}^V({\bf \Delta})&=&\frac{{\bf \Delta}+i[{\bf e}^*{\bf
\Delta}]}{2m_a}
F_1({\bf \Delta}),\\
\label{48}
J_0^A({\bf \Delta})&=&\frac{({\bf e}^*{\bf \Delta})}{2m_a}F_1({\bf
\Delta}),\\
\label{49}
{\bf J}^A({\bf \Delta})&=&{\bf e}^*F_2(\Delta).
\end{eqnarray}
Functions $F_1$ and $F_2$ arise from the lower and the upper
components of
Dirac spinors
\begin{equation}
\label{50}
u^{\lambda}_a(p)=\left(\frac{\varepsilon_a(p)+m_a}{2\varepsilon_a(p)}
\right)^{1/2}\left(
\begin{array}{c}
1\\
\frac{\vec \sigma {\bf p}}{\varepsilon_a(p)+m_a}
\end{array}
\right)\chi^{\lambda},
\end{equation}
and are equal to
\begin{eqnarray}
\label{51}
F_1({\bf \Delta})&=&\frac{2m_a}{\varepsilon_a({\bf q}+{\bf
\Delta})+m_a}
\left(\frac{\varepsilon_a({\bf q}+{\bf
\Delta})+m_a}{2\varepsilon_a({\bf q}
+{\bf \Delta})}\right)^{1/2}\sqrt{\frac{E_A}{M_A}},\\
\label{52}
F_2({\bf \Delta})&=&\left(\frac{\varepsilon_a({\bf q}+{\bf
\Delta})+m_a}
{2\varepsilon_a({\bf q}+{\bf
\Delta})}\right)^{1/2}\sqrt{\frac{E_A}{M_A}}.
\end{eqnarray}
Near $q^2=q^2_{max}$ it can be written as
\begin{eqnarray}
\label{53}
F_1({\bf \Delta})&=&\frac{\sqrt{2}(1+{\bf
\Delta}^2/M_A^2)^{1/2}}{\left({1+{\bf \Delta}^2/m_a^2+
\sqrt{1+{\bf \Delta}^2/m_a^2}}\right)^{1/2}},\\
\label{54}
F_2({\bf \Delta})&=&\frac{1}{\sqrt{2}}\left(1+
\frac{1}{\sqrt{1+{\bf \Delta}^2/m_a^2}} \right)^{1/2}
\left(1+\frac{{\bf \Delta}^2}{M_A^2}\right)^{1/2}.
\end{eqnarray}

The dependence of the formfactors on the momentum transfer is fixed by
extrapolating their behavior near $q^2=q^2_{max}$ (${\bf \Delta}=0$)
point
over the kinematically allowed region
\begin{eqnarray}
\label{55}
f_+({\bf \Delta})& = &f_+(0)I({\bf \Delta})F_1({\bf \Delta}),\\
A_1({\bf \Delta})& = &A_1(0)I({\bf \Delta})F_2({\bf \Delta}),\\
A_2({\bf \Delta})& = &A_2(0)I({\bf \Delta})F_1({\bf \Delta}),\\
V({\bf \Delta})& = &V(0)I({\bf \Delta})F_1({\bf \Delta}),
\end{eqnarray}
where
\begin{equation}
\label{59}
I({\bf \Delta})=\int\frac{d^3{\bf p}}{(2\pi)^3}\bar\Psi_A\left(
{\bf p}+\frac{2\varepsilon_q}{E_A+M_A}{\bf \Delta}\right)\Psi_B({\bf
p}).
\end{equation}
Introducing the variable
\begin{equation}
\label{60}
w\equiv v_Av_B=\frac{M_B^2+M_A^2-q^2}{2M_AM_B},
\end{equation}
where $v_A$ and $v_B$ are meson velocities, and taking into account
that
\begin{equation}
\label{61}
{\bf \Delta}^2=({\bf p}_B-{\bf p}_A)^2=\frac{(M_B^2+M_A^2-q^2)^2}
{4M_B^2}-M_A^2=M_A^2(w^2-1),
\end{equation}
we can rewrite (\ref{55})--(47) in the form
\begin{eqnarray}
\label{62}
f_+(w)& = &f_+(1)I(w)\left(\frac{2}{w+1}\right)^{1/2},\\
A_1(w)& = &A_1(1)I(w)\left(\frac{w+1}{2}\right)^{1/2},\\
A_2(w)& = &A_2(1)I(w)\left(\frac{2}{w+1}\right)^{1/2},\\
V(w)& = &V(1)I(w)\left(\frac{2}{w+1}\right)^{1/2},
\end{eqnarray}
Substitution of the Gaussian wave functions (\ref{29}) in (\ref{59})
results
in
\begin{equation}
\label{66}
I(w)=\exp{\left(-\frac{2\bar \Lambda^2}{\beta^2_{M_A}+\beta^2_{M_B}}
\frac{w-1}{w+1}\right)}I(1),
\end{equation}
where $\bar \Lambda=<\varepsilon_q>$ is a mean value of light quark
energy inside
meson. In our model $\bar \Lambda$ corresponds to HQET parameter
$\bar \Lambda=M-m_Q$, which determines the energy carried by light
degrees
of freedom, and is found to be~\cite{9}: $\bar \Lambda=0.54$~GeV.

In the limit of infinitely heavy $b$ and $a$ quarks the $w$
dependence of
eqs.~(\ref{62})--(\ref{66}) is determined by
the Isgur-Wise function of our model
\begin{equation}
\label{IW}
\xi(w)=\left(\frac{2}{w+1}\right)^{1/2}\exp\left(-\frac{\bar
\Lambda^2}
{\beta^2}\frac{w-1}{w+1}\right)
\end{equation}
and the ratios of form factors  satisfy all
constraints imposed by HQET~\cite{7}.

Using (\ref{31})--(\ref{34}) and (\ref{35})--(\ref{45}) we have
calculated form factors for
$B\to D(D^*)l\nu_l$, $D\to K(K^*)l\nu_l$ and $D_s\to\varphi l\nu_l$
exclusive decays. The results obtained in our model for $D\to
K(K^*)l\nu_l$
are compared with appropriate experimental data and various model
predictions in Table~1. New values of form factors for $D\to
K(K^*)l\nu_l$
are somewhat larger then our previous results~\cite{14} because of
more consistent
relativistic treatment of $s$ quark and slight change in the value of
the mixing coefficient $\varepsilon$. Note that while the other
potential models agree with the experimental determination of $V(0)$,
but
fail to predict $A_1(0)$ and $A_2(0)$, our model predicts correct
values of
$A_1(0)$ and $A_2(0)$, but gives too low value of $V(0)$. The reason
for that
is not clear. The contribution from
form factor $V(0)$ in the total width is kinematicaly suppresed. So,
despite the
above mentioned descripancy, we have got the $D\to K^*l\nu_l$ width
in accord with experimental data.

The ratios of form factors $R_2=A_2(0)/A_1(0)$ and
$R_V=V(0)/A_1(0)$ are given in Table~2.

The obtained branching ratios are\\
\centerline{$B(D^0\to K^{*^-}e^+\nu_e)=1.9\%$,  for $\tau_{D^0}=
0.415\times 10^{-12}$s;}
\centerline{$B(D^+\to \bar K^{*^0}e^+\nu_e)=4.9\%$,
 for $\tau_{D^+}=1.060\times 10^{-12}$s;}
to be compared with the experimental average data~\cite{22}\\
\centerline{$B^{exp}(D^0\to K^{*^-}e^+\nu_e)=(2.0\pm0.4)\%$,}
\centerline{$B^{exp}(D^+\to \bar K^{*^0}e^+\nu_e)=(4.8\pm0.4)\%$.}
The ratio $\Gamma(D\to K^*e\nu_e)/\Gamma(D\to Ke\nu_e)$ and the ratio
of the
longitudinal and transverse decay widths $\Gamma_L/\Gamma_T$ also
agree well
with experiment~(see Table~3).

For $D_s\to\varphi l\nu_l$ decay form factors our predictions are\\
\centerline{$A_1(0)=0.63$, $A_2(0)=0.35$ and $V(0)=1.06$.}
As experiment provides us with $R_2$ and $R_V$ ratios for
$D_s\to\varphi l\nu_l$, we compare it with predicted values in
Table~4.
It is not clear why experimental ratio $R_2$ for $D_s\to\varphi
l\nu_l$
so differs from that of $D\to\ K^* l\nu_l$. In RQM we get
approximately
equal ratios $R_2$ for both decays, because general structure and the
signs
of the potential-dependent corrections in (\ref{31})--(\ref{34}) are
almost
the same. It can be expected that the experimental results for
$D_s\to\varphi l\nu_l$ form factor ratios will change in the future.
Anyway,
the experimental uncertainties are still too large to conclude that
there is
a serious descripancy between RQM and the experimental data in this
case.

For $D_s\to\varphi l\nu_l$ branching ratio we have
$B(D_s\to\varphi l\nu_l)=2.5\%$, while experiment gives $B^{exp}
(D_s\to\varphi l\nu_l)=(1.88\pm0.29)\%$~\cite{22}.

In $B\to D^*l\nu_l$ decay, since both $b$ and $c$ quarks are heavy,
relativistic corrections are not so significant, but the Lorentz-
structure of
quark-antiquark potential has an important influence on the values of
form factors. We have found our results for $R_2$ and $R_V$ to be in
a good
agreement with the experimental data~\cite{23} and HQET-based
predictions~\cite{25}.
Measurments and predictions for the ratios of the form factors for
$B\to D^*l\nu_l$, evaluated at $q^2=q^2_{max}$, are shown in Table~5.
We have obtained the following results for $B\to D^*l\nu_l$ and
$B\to Dl\nu_l$ branching ratios\\
\centerline{$B(B\to D^*l\nu_l)=33.8\times|V_{bc}|^2$, $B(B\to
Dl\nu_l)=
19.8\times|V_{bc}|^2$, for $\tau_{B^0}=1.5\times10^{-12}$s.}\\
It should be compared to the experimental data\\
\centerline{$B(B^0\to D^-e^+\nu_e)=(2.0\pm0.7\pm0.6)\%$
ARGUS~\cite{26}}
\centerline{$B(B^0\to D^-e^+\nu_e)=(1.8\pm0.6\pm0.3)\%$ CLEO-
I~\cite{27}}
\centerline{$B(B^0\to D^{*-}e^+\nu_e)=(4.7\pm0.5\pm0.5)\%$
ARGUS~\cite{28}}
\centerline{$B(B^0\to D^{*-}e^+\nu_e)=(4.0\pm0.4\pm0.6)\%$ CLEO-
I~\cite{27}.}
As a result we can extract the value of CKM matrix element $V_{cb}$\\
\centerline{$|V_{cb}|=0.036\pm0.004$.}

\subsection{Differential Distributions}
The differential decay rate~\cite{GS} can be expressed in terms of
two dimensionless
variables $x=E_l/M_B$ and $w=v_Av_B$, where $E_l$ is the lepton energy
\begin{eqnarray}
\label{67}
\nonumber
\frac{d^2\Gamma}{dxdw}&=&\frac{|V_{ab}|^2G_F^2M_B^5}{32\pi^3}
\bigg[\Big(\frac{2M_A}{M_B}w-1-\rho\Big)\Big(W_1(w)-2W_3(w)
\big(1-2x-\frac{M_A}{M_B}w\big)\Big)\\
&&+2W_2(w)\Big(\rho+(1-2x)^2
-\frac{2M_A}{M_B}w(1-2x)\Big)\bigg],
\end{eqnarray}
Here $W_{1,2,3}(w)$ are connected with semileptonic form factors

a) For $0^-\to 0^-$ transition
\begin{eqnarray}
\label{68}
W_1(w)&=&0,\\
\label{69}
W_2(w)&=&\frac{2M_A}{M_B}|f_+(w)|^2,\\
\label{70}
W_3(w)&=&0,
\end{eqnarray}

b) For $0^-\to 1^-$ transition
\begin{eqnarray}
\label{71}
W_1(w)&=&\frac{2M_A}{M_B}\left(1+\frac{M_A}{M_B}\right)^2A_1^2(w)+
\frac{2M_A}{M_B}\frac{4M_A^2}{(M_A+M_B)^2}V^2(w)(w^2-1),\\
\label{72}
\nonumber
W_2(w)&=&\frac{M_B}{2M_A}\left(1+\frac{M_A}{M_B}\right)^2A_1^2(w)-
\frac{2M_AM_B}{M_A+M_B}V^2(w)\left(1+\rho-\frac{2M_A}{M_B}w\right)\\
&&+2A_1(w)A_2(w)\left(\frac{M_A}{M_B}-
w\right)+\frac{2M_AM_B}{(M_A+M_B)^2}
A_2^2(w)(w^2-1),\\
\label{73}
W_3(w)&=&\frac{4M_A}{M_B}A_1(w)V(w).
\end{eqnarray}
The kinematiclly allowed region is presented in Fig.~3 where the
lower bound
curve $w_m(x)$ has the following shape
\begin{equation}
\label{74}
w_m(x)=\frac{M_B}{2M_A}(1-2x)+\frac{M_A}{2M_B}\frac{1}{1-2x}.
\end{equation}
The analytical expression for $d\Gamma/dx$ distribution depends on
$q^2$
behavior of form factors. Using (\ref{62})--(\ref{66}), we obtain
after
the integration over $w$
\begin{equation}
\label{75}
\frac{d\Gamma}{dx}=\frac{G_F^2|V_{cb}|^2M_B^5}{32\pi^3}\bigg[
e^{-\alpha\frac{w_m(x)-1}{w_m(x)+1}}K_1(x)+\sinh{(\alpha\chi_-(x))}
e^{-\alpha\chi_+(x)}K_2(x)+K_3(x)\int_{w_m(x)}^{w_0}\!\!\!\!\!\!\!
dw\,e^{-\alpha{\frac{w-1}{w+1}}}\bigg],
\end{equation}
where
\begin{eqnarray}
\label{76}
\chi_-(x)&=&\frac{4M_AM_B^3}{(M_A+M_B)^4}\frac{x(1-R)}{1-\frac{2M_B^2}
{(M_A+M_B)^2}x(1-R)},\\
\label{77}
\chi_+(x)&=&\left(\frac{M_B-M_A}{M_B+M_A}\right)^2\frac{1-
\frac{M_A^2+M_B^2}
{M_B^2-M_A^2}\frac{2M_B^2}{M_B^2-M_A^2}x(1-R)}{1-
\frac{2M_B^2}{M_B^2-M_A^2}x(1-R)},
\end{eqnarray}
$$
R=\frac{\rho}{1-2x},\quad \rho=\frac{M_A^2}{M_B^2},\quad
\alpha=\frac{4\bar \Lambda^2}
{\beta_{M_A}^2+\beta_{M_B}^2},\quad w_0=\frac{M_B^2+M_A^2}{2M_AM_B}.
$$
Functions $K_{1,2,3}(x)$ take different forms for $0^-\to 0^-$ and
$0^-\to 1^-$
decays and are given in Appedix B.

The $(1/\Gamma)(d\Gamma/dx)$ distributions for $B\to D(D^*)l\nu_l$,
$D\to
K(K^*)l\nu_l$ and $D_s\to\varphi l\nu_l$ decays are shown in Figs.~4--
6. All
curves are normalised by the corresponding decay width $\Gamma$, i.e.
the area
under each curve is equal to $1$.

\section{Conclusion}
Using the quasipotential approach, we have obtained the expressions
for
semileptonic decay form factors with the consistent account of
relativistic effects.
This account includes more careful
consideration of the $s$ quark  contribution than it was in our
previous work~
\cite{14} and
results in the small shift in the values of $D\to K(K^*)l\nu_l$ decay
form
factors, which are in a good agreement with
measurments.  Our model provides more accurate values for the ratio
$A_2(0)/A_1(0)$ in the $D\to K(K^*)l\nu_l$ decay and for the ratio
$\Gamma(D\to K^*l\nu_l)/\Gamma(D\to Kl\nu_l)$ in comparison with
the other models. We have also calculated form factors and branching
ratios
for $B\to D(D^*)l\nu_l$, $D\to K(K^*)l\nu_l$ and $D_s\to\varphi
l\nu_l$.
The extracted value of $|V_{cb}|=0.036\pm0.003$ is lower than the
previous
one~\cite{14} because of the changes in the form factors as well as
in the
$B$ meson lifetime. We should emphasize that in order to get reliable
results
for $D$ meson semileptonic decays it is necessary to take into
consideration
all possible relativistic effects, including the transformation of
meson
wave function~(\ref{28}) from the rest frame, which is ignored in many
quark models.

The proposed $q^2$ dependence of the form factors is
used for the determination of differential semileptonic distributions
in the
case of pseudoscalar and vector final states.

It should be noted that obtained expressions for semileptonic form
factors are
valid for all $B$ and $D$ meson decays, except the decays into mesons
containing two light quarks ($\pi$, $\rho$ mesons), where one cannot
expand
in either ${\bf p}^2/m^2$ or ${\bf p}^2/\varepsilon^2$ at
$q^2=q^2_{max}$
point in the vertex function~(\ref{27}). The solution of this problem
is proposed
in~\cite{35}.

The analysis has shown that Lorentz-structure of quark-antiquark
potential
plays an important role in heavy meson semileptonic decays. We have
got
experimentally motivated and HQET based arguments to conclude that the
confining potential has predominantly Lorentz-vector (with Pauli-term)
structure $\varepsilon=\,-1$. Assuming also long-range anomalous
chromomagnetic
moment of quark to be $\kappa=\,-1$ we have obtained satisfactory
description
of all considered $B$ and $D$ semileptonic decays.

We argue that small number of parameters, most of
which were fixed previously, and an agreement with HQET for the
structure of
leading,
subleading and second order terms in $1/m_Q$ expansion make RQM a
reliable tool
for the investigation of heavy meson physics.

In this paper we have restricted ourselves with  $0^-\to 0^-$
and $0^-\to 1^-$ transitions. One more practically important case of
the
decay into $P$~wave final state (i.e. $B\to D^{**}l\nu_l$) will be
considered
in the future.

\vskip 1cm
{\Large \bf Acknowledgments}
\vskip 4mm
We would like to thank B.A.Arbuzov, M.Beyer, A.G.Grozin,
J.G.K\"orner, T.Mannel,
V.A.Matveev, M.Neubert, V.I.Savrin, and A.I.Studenikin for the
interest in our
work and helpful discussions. We are also appreciate the help of
G.G.Likhachev in preparation of this paper for publication and
wish to thank the Interregional Centre for Advanced Studies (Moscow)
for the kind hospitality during all stages of this work. This work was
supported in part by the Russian Foundation for Fundamental Research
under
Grant No.~94-02-03300-a.
\bigskip
\newpage
\rm
\noindent {\Large\bf Appendix A. Exclusive Semileptonic Decay Form
Factors
at $q^2=q^2_{max}$ point}

{\bf arXiv note - added Dec99 - TeX source for this appendix was corrupt
on initial submission, changed to verbatim environment so that the 
rest of the paper can be processed.}

\begin{verbatim}
\medskip
\begin{eqnarray}
\label{35}
\nonumber
f_+^{(1)}(\qm)&=&\sqrt{\frac{M_A}{M_B}}\int\frac{d^3{\bf p}}{(2\pi)^3}
\bar\Psi_A({\bf
p})\left(\frac{\varepsilon_a+m_a}{2\varepsilon_a}\right)^{1/2}
\Bigg[1+\frac{M_B-M_A}{\varepsilon_a+m_a}
-\frac{{\bf p}^2}{8}\bigg(\frac{1}{m_b^2}
-\frac{4}{m_b(\varepsilon_a+m_a)}\\
\nonumber
&&+\frac{M_B-
M_A}{\varepsilon_a+m_a}\Big(\frac{1}{m_b^2}+\frac{4}{\varepsilon_a
(\varepsilon_a+m_a)}+\frac{2}{3\varepsilon_am_a}\Big)+\frac{4}{3}\frac
\nonumber
&&\times\Big(\frac{1}{\varepsilon_a+m_a}-
\frac{1}{2m_b}\Big)\frac{\varepsilon_a
-\varepsilon_q}{\varepsilon_a \varepsilon_q} \bigg)
+\frac{M_B-M_A}{3}\bigg(\frac{1}{\varepsilon_a+m_a}+
\frac{1}{2m_b}\\
&&-\frac{{\bf p}^2}{8m_b^2}\Big(
\frac{1}{\varepsilon_a+m_a}+\frac{3}{2m_b}
\Big)\Bigg)\frac{\varepsilon_q}{M_A}({\bf p}\overleftarrow
{\frac{\partial}{\partial{\bf p}}})\Bigg]\Psi_B({\bf p}),\\
\label{36}
\nonumber
f_{+S}^{(2)}(\qm)&=&\sqrt{\frac{M_A}{M_B}}\int\frac{d^3{\bf
p}}{(2\pi)^3}
\bar\Psi_A({\bf
p})\left(\frac{\varepsilon_a+m_a}{2\varepsilon_a}\right)^{1/2}
\Bigg[-\frac{M_B-M_A}{2\varepsilon_a}\frac{M_B-\varepsilon_b-
\varepsilon_q}
{\varepsilon_a}+
\frac{M_B-M_A}{12\varepsilon_a}{\bf p}^2\\
\nonumber
&&\times\Big(\frac{1}{\varepsilon_a^2}+\frac{1}{m_b^2}\Big)-
\frac{M_B-M_A}{12}\Big(\frac{1}{\varepsilon_a^2}+\frac{1}{m_b^2}\Big)
\big(M_B+M_A-\varepsilon_b-\varepsilon_a-2\varepsilon_q\big)\\
&&\times\frac{\varepsilon_q}{M_A}({\bf p}\overleftarrow
{\frac{\partial}{\partial{\bf p}}})\Bigg]\Psi_B({\bf p}),\\
\label{37}
\nonumber
f_{+V}^{(2)}(\qm)&=&\sqrt{\frac{M_A}{M_B}}\int\frac{d^3{\bf
p}}{(2\pi)^3}
\bar\Psi_A({\bf
p})\left(\frac{\varepsilon_a+m_a}{2\varepsilon_a}\right)^{1/2}
\Bigg[\frac{M_B-M_A}{\varepsilon_a}\frac{{\bf
p}^2}{12}\bigg((1+\kappa)
\Big(\frac{1}{\varepsilon_a^2}-\frac{1}{m_b^2}\Big)\\
\nonumber
&&-
\frac{1}{\varepsilon_q}\Big(\frac{2}{\varepsilon_a+m_a}+\frac{1}{m_b}
\Big)\bigg)+
\frac{M_B-M_A}{12}(1+\kappa)\Big(\frac{1}{\varepsilon_a^2}-
\frac{1}{m_b^2}
\Big)\big(M_B-M_A-\varepsilon_q+\varepsilon_a\big)\\
\nonumber
&&\times\frac{\varepsilon_q}{M_A}({\bf p}\overleftarrow
{\frac{\partial}{\partial{\bf p}}})
+\frac{M_B-M_A}{6\varepsilon_q}\Big(\frac{1}{\varepsilon_a+m_a}
+\frac{1}{2m_b}\Big)\big(M_B+M_A-\varepsilon_b-\varepsilon_a-
2\varepsilon_q\big)\\
&&\times\frac{\varepsilon_q}{M_A}({\bf p}\overleftarrow
{\frac{\partial}{\partial{\bf p}}})\Bigg]\Psi_B({\bf p}),\\
\nonumber
A_1^{(1)}(\qm)&=&(M_A+M_B)\sqrt{4M_AM_B}\int\frac{d^3{\bf
p}}{(2\pi)^3}
\bar\Psi_A({\bf
p})\left(\frac{\varepsilon_a+m_a}{2\varepsilon_a}\right)^{1/2}
\bigg(1-\frac{{\bf p}^2}{8}\Big(\frac{1}{m_b^2}\\
&&+\frac{4}{3(\varepsilon_a+m_a)m_b}\Big)\bigg)\Psi_B({\bf p}),\\
\label{39}
A_{1S}^{(2)}(\qm)&=&A_{1V}^{(2)}(\qm)=0,\\
\label{40}
\nonumber
A_2^{(1)}(\qm)&=&\frac{1}{(M_A+M_B)\sqrt{4M_AM_B}}\int\frac{d^3{\bf
p}}
{(2\pi)^3}\bar\Psi_A({\bf
p})\left(\frac{\varepsilon_a+m_a}{2\varepsilon_a}
\right)^{1/2}\Bigg[\bigg(1+\frac{M_A}{M_B}\bigg)\\
\nonumber
&&\times\bigg(1
-\frac{{\bf
p}^2}{2}\Big(\frac{1}{4m_b^2}+\frac{1}{3m_b(\varepsilon_a+m_a)}
\Big)\bigg)-
\frac{2M_A^2}{M_B(\varepsilon_a+m_a)}\bigg(1-\frac{{\bf p}^2}{8}
\Big(\frac{1}{m_b^2}\\
\nonumber
&&+\frac{4}{\varepsilon_a(\varepsilon_a+m_a)}+
\frac{2}{3\varepsilon_am_b}\Big)
-\frac{{\bf p}^2}{6M_A\varepsilon_a}
\Big(1+\frac{\varepsilon_a+m_a}{2m_b}\Big)\bigg)\\
\nonumber
&&-\frac{2M_A\varepsilon_q}{(\varepsilon_a+m_a)M_B}\bigg(
\frac{1}{3}\Big(1+\frac{\varepsilon_a+m_a}{2m_b}
-\frac{{\bf p}^2}{8m_b^2}
\Big(1+\frac{3(\varepsilon_a+m_a)}{2m_b}\Big)\Big)\\
&&+\frac{M_B-M_A}{3m_b}\Big(1-\frac{3{\bf p}^2}{8m_b^2}\Big)\bigg)
({\bf p}\overleftarrow
{\frac{\partial}{\partial{\bf p}}})\Bigg]\Psi_B({\bf p}),\\
\label{41}
\nonumber
 A_{2S}^{(2)}(\qm)&=&\frac{1}{(M_A+M_B)\sqrt{4M_AM_B}}\int\frac{d^3{\bf
{(2\pi)^3}\bar\Psi_A({\bf
p})\left(\frac{\varepsilon_a+m_a}{2\varepsilon_a}
\right)^{1/2}\frac{M_A^2}{\varepsilon_aM_B}\Bigg[-\frac{M_B-
\varepsilon_b
-\varepsilon_q}{\varepsilon_a}\\
\nonumber
&&+\frac{{\bf
p}^2}{6}\Big(\frac{1}{\varepsilon_a^2}+\frac{1}{m_b^2}\Big)-
\frac{\varepsilon_a}{6}\big(M_B+M_A-\varepsilon_b-\varepsilon_a-
2\varepsilon_q\big)
\Big(\frac{1}{\varepsilon_a^2}+\frac{1}{m_b^2}\Big)\frac{\varepsilon_q
{M_A}({\bf p}\overleftarrow
{\frac{\partial}{\partial{\bf p}}})\Bigg]\Psi_B({\bf p}),\\
\\
\label{42}
\nonumber
A_{2V}^{(2)}(\qm)&=&\frac{1}{(M_A+M_B)\sqrt{4M_AM_B}}\int\frac{d^3{\bf
{(2\pi)^3}\bar\Psi_A({\bf
p})\left(\frac{\varepsilon_a+m_a}{2\varepsilon_a}
\right)^{1/2}\Bigg[\frac{M_A^2}{\varepsilon_aM_B}\bigg(\frac{{\bf
p}^2}{6}
(1+\kappa)\\
\nonumber
&&\times\Big(\frac{1}{\varepsilon_a^2}-\frac{1}{m_b^2}\Big)+
\frac{\varepsilon_a}{6}\big(M_B-M_A-\varepsilon_b+\varepsilon_a\big)
(1+\kappa)\Big(\frac{1}{\varepsilon_a^2}-\frac{1}{m_b^2}\Big)
\frac{\varepsilon_q}{M_A}({\bf p}\overleftarrow
{\frac{\partial}{\partial{\bf p}}})\bigg)\\
&&-\frac{{\bf p}^2}{6\varepsilon_q}
\nonumber
\Big(\frac{2}{\varepsilon_a+m_a}
+\frac{1}{m_b}\Big)+
\frac{\varepsilon_a}{6M_A}\Big(\frac{2}{\varepsilon_a+m_a}+\frac{1}{m_
&&\times\big(M_B+M_A-\varepsilon_b-\varepsilon_a-2\varepsilon_q\big)
({\bf p}\overleftarrow
{\frac{\partial}{\partial{\bf p}}})\Bigg]\Psi_B({\bf p}),\\
\label{43}
\nonumber
V^{(1)}(q^2_{max})&=&\frac{1}{M_A+M_B}\sqrt{\frac{M_A}{M_B}}\int\frac{
{(2\pi)^3}\bar\Psi_A({\bf
p})\left(\frac{\varepsilon_a+m_a}{2\varepsilon_a}
\right)^{1/2}\frac{1}{\varepsilon_a+m_a}\Bigg[1-\frac{{\bf p}^2}{8}
\Big(\frac{1}{m_b^2}\\
\nonumber
&&+\frac{4}{\varepsilon_a(m_a+\varepsilon_a)}
-\frac{2}{3\varepsilon_am_b}\Big)-
\frac{{\bf p}^2}{12M_A}\Big(1+\frac{\varepsilon_a+m_a}{2m_b}\Big)
\frac{\varepsilon_a+\varepsilon_q}{\varepsilon_a\varepsilon_q}-
\frac{{\bf p}^2}{12M_A\varepsilon_a}\Big(1-
\frac{\varepsilon_a+m_a}{2m_b}\Big)\\
&&+\frac{1}{3}\Big(1-\frac{\varepsilon_a+m_a}{2m_b}-\frac{{\bf
p}^2}{8m_b^2}
\Big(1+\frac{3(\varepsilon_a+m_a)}{2m_b}\Big)\Big)
\frac{\varepsilon_q}{M_A}({\bf p}\overleftarrow
{\frac{\partial}{\partial{\bf p}}})\Bigg]\Psi_B({\bf p}),\\
\label{44}
\nonumber
V^{(2)}_S(q^2_{max})&=&\frac{1}{M_A+M_B}\sqrt{\frac{M_A}{M_B}}\int\fra
{(2\pi)^3}\bar\Psi_A({\bf
p})\left(\frac{\varepsilon_a+m_a}{2\varepsilon_a}
\right)^{1/2}\frac{1}{2\varepsilon_a}\Bigg[-\frac{M_B-\varepsilon_b
-\varepsilon_q}{\varepsilon_a}\\
\nonumber
&&+\frac{{\bf p}^2}{6}\Big(\frac{1}{\varepsilon_a^2}-
\frac{1}{m_b^2}\Big)-
\frac{\varepsilon_a}{6}\big(M_B+M_A-\varepsilon_b-\varepsilon_a-
2\varepsilon_q\big)\\
&&\times\Big(\frac{1}{\varepsilon_a^2}-
\frac{1}{m_b^2}\Big)\frac{\varepsilon_q}
{M_A}({\bf p}\overleftarrow
{\frac{\partial}{\partial{\bf p}}})\Bigg]\Psi_B({\bf p}),\\
\label{45}
\nonumber
V^{(2)}_V(q^2_{max})&=&\frac{1}{M_A+M_B}\sqrt{\frac{M_A}{M_B}}\int\fra
{(2\pi)^3}\bar\Psi_A({\bf
p})\left(\frac{\varepsilon_a+m_a}{2\varepsilon_a}
\right)^{1/2}\frac{1}{2\varepsilon_a}\Bigg[\Big(\frac{{\bf p}^2}{6}
+\frac{\varepsilon_a}{6}\\
\nonumber
&&\times\big(M_B-M_A-\varepsilon_b+\varepsilon_a\big)
\times\frac{\varepsilon_q}{M_A}({\bf
p}\overleftarrow{\frac{\partial}{\partial
{\bf p}}})\Big)
(1+\kappa)\Big(\frac{1}{\varepsilon_a^2}+\frac{1}{m_b^2}-
\frac{1}{\varepsilon_q}\Big(\frac{2}{\varepsilon_a+m_a}+
\frac{1}{m_b}\Big)\Big)\\
\nonumber
&&-\frac{{\bf
p}^2\varepsilon_a}{6\varepsilon_q}\Big(\frac{2}{\varepsilon_a+m_a}
-\frac{1}{m_b}\Big)
+\frac{\varepsilon_a}{6M_A}
\big(M_B+M_A-\varepsilon_b-\varepsilon_a-2\varepsilon_q\big)\\
&&\times\Big(\frac{2}{\varepsilon_a+m_a}-\frac{1}{m_b}\Big)({\bf
p}\overleftarrow
{\frac{\partial}{\partial{\bf p}}})\Bigg]\Psi_B({\bf p}),
\end{eqnarray}
 here $({\bf p}\overleftarrow{\partial/\partial{\bf p}})$ acts
to the left on the wave function $\bar\Psi_A({\bf p})$.
In the limit ${\bf p}^2/m^2\to 0$ the above form factors reduce to the
standard expressions, obtained in the nonrelativistic quark models.
\end{verbatim}
\vskip 1cm

\noindent {\Large\bf Appendix B. Functions $K_{1,2,3}(x)$ for
$(1/\Gamma)
d\Gamma/dx$ Differential Distributions}

\medskip

a) In the case of $0^-\to 0^-$ transition
\begin{eqnarray}
\label{78}
K_1(x)&=&\frac{2M_A}{M_B}x(1-2x)(1-R)|f_+(1)|^2,\\
\label{79}
K_2(x)&=&\frac{2M_A}{M_B}\left(1+\frac{M_A}{M_B}\right)^2(1-
2x)|f_+(1)|^2,\\
\label{80}
K_3(x)&=&\frac{M_A}{M_B}\left(\rho+(1-2x)^2+\frac{2M_A}{M_B}(1-2x)
-\frac{4M_A}{M_B}\alpha(1-2x)\right)|f_+(1)|^2.
\end{eqnarray}

b) In the case of $0^-\to1^-$ transition
\begin{eqnarray}
\label{81}
\nonumber
K_1(x)&=&\frac{M_B}{M_A}x(1-R)\big(G_1(x)+\alpha
G_2(x)+\frac{2}{3}\alpha^2
G_3(x)\big)+\frac{1}{2}\big(G_2(x)+\frac{2}{3}\alpha G_3(x)\big)\\
\nonumber
&&\times\Big(x(1-r)\big(1+\frac{M_B}{M_A}\big)^2-
\frac{M_B^2}{M_A^2}x^2(1-R)^2\Big)
+\frac{1}{3}G_3(x)\Big(\frac{3M_B}{M_A}x(1-
R)\frac{(M_A+M_B)^4}{4M_A^2M_B^2}\\
&&+\frac{3M_B^2}{M_A^2}\Big(\frac{M_A+M_B}{2M_AM_B}\Big)^2x^2(1-R)^2
-\Big(\frac{M_A+M_B}{8M_A^3M_B^3}\Big)^6x^3(1-R)^3\Big),\\
\label{82}
\nonumber
K_2(x)&=&-\frac{(M_A+M_B)^2}{M_AM_B}\big(G_1(x)+\alpha
G_2(x)+\frac{2}{3}
\alpha^2G_3(x)\big)\\
&&-\frac{(M_A+M_B)^4}{M_A^2M_B^2}\big(G_2(x)+\frac{2}{3}
\alpha G_3(x)\big)-\frac{(M_A+M_B)^6}{12M_A^3M_B^3}G_3(x),\\
\label{83}
K_3(x)&=&-G_4(x)-2\alpha G_1(x)-2\alpha^2G_2(x)-
\frac{4}{3}\alpha^3G_3(x),
\end{eqnarray}
where $G_{1,2,3,4}(x)$ depend on the values of form factors at $w=1$
point
\begin{eqnarray}
\label{84}
\nonumber
G_1(x)&=&V^2(1)\bigg(\frac{16M_A^2}{(M_A+M_B)^2}\Big(1-
2x+\frac{M_A}{M_B}\Big)^2
-\frac{16M_A^2}{M_B^2}\Big(1-2x+\frac{2M_A}{M_B}\Big)\bigg)\\
\nonumber
&&+\frac{16M_AM_B}{(M_A+M_B)^2}A_2^2(1)\Big(1-
2x+\frac{M_A}{M_B}\Big)^2-
\frac{8M_A}{M_B}A_1(1)V(1)\Big(1+\frac{M_A}{M_B}\Big)^2\Big(1-2x+
\frac{M_A}{M_B}\Big)\\
&&-\frac{4M_A}{M_B}A_1(1)A_2(1)\Big(1+\frac{M_B}{M_A}\Big)
\Big(1-2x+\frac{M_A}{M_B}\Big)^2,\\
\label{85}
\nonumber
G_2(x)&=&\frac{16M_A^2}{(M_A+M_B)^2}\frac{M_A}{M_B}V^2(1)\Big(2(1-
2x)+\big(
1+\frac{M_A}{M_B}\big)^2+\frac{4M_A}{M_B}\Big)-
\frac{8M_AM_B}{(M_A+M_B)^2}A_2^2(1)\\
\nonumber
&&\times\Big(\big(1-2x+\frac{M_A}{M_B}\big)^2
+\frac{4M_A}{M_B}(1-
2x)\Big)+\frac{M_A}{M_B}A_1^2(1)\big(1+\frac{M_A}{M_B}
\big)^2\Big(\big(1+\frac{M_A}{M_B}\big)^2\\
\nonumber
&&\times\frac{M_B^2}{2M_A^2}\big(1-2x+\frac{M_A}{M_B}\big)^2\Big)+
\frac{8M_A^2}{M_B^2}A_1(1)V(1)\Big(
2(1-2x)+\big(1+\frac{M_A}{M_B}\big)^2+\frac{2M_A}{M_B}\Big)\\
&&+4A_1(1)A_2(1)\big(1-2x+\frac{M_A}{M_B}\big)^2+
\frac{8M_A}{M_B}A_2(1)A_1(1)(1-2x)
\big(1+\frac{M_A}{M_B}\big),\\
\label{86}
\nonumber
G_3(x)&=&-\frac{8M_A}{M_B}A_1(1)A_2(1)(1-2x)-
\frac{16M_A^3}{M_B^3}A_1(1)V(1)+
\frac{16M_A^2}{(M_A+M_B)^2}\Big(A_2^2(1)(1-2x)\\
&&-\frac{2M_A^2}{M_B^2}V^2(1)\Big)+\frac{2M_A^2}{M_B^2}
\Big(1+\frac{M_A}{M_B}\Big)^2A_1^2(1)\Big(-1+\frac{M_B^2}{2M_A^2}(1-
2x)
\Big),\\
\label{87}
G_4(x)&=&\frac{8M_AM_B}{(M_A+M_B)^2}V^2(1)\Big(1+\frac{M_A}{M_B}\Big)^
\Big(1-2x+\frac{M_A}{M_B}\Big)^2.
\end{eqnarray}

\newpage
\noindent

Table~1. Theoretical predictions and experimental data for form
factors
in $D\to Kl\nu_l$ and $D\to D^*l\nu_l$.

\medskip
\begin{tabular}{lcccc}
\hline
\hline
Ref.& $f_+(0)$ & $V_0(0)$ & $A_1(0)$ & $A_2(0)$\\
\hline
 Exp. Average~\cite{22} & $0.75\pm0.02\pm0.02$ & $1.1\pm0.2$ &
$0.56\pm0.04$ & $0.40\pm0.08$\\
\hline
 Theory & & & & \\
 RQM & $0.73$ & $0.62$ & $0.63$ & $0.43$\\
 ISGW \cite{1} & $0.82$ & $1.1$ & $0.8$ & $0.8$\\
 BSW \cite{2} & $0.76$ & $1.3$ & $0.88$ & $1.2$\\
 AW \cite{3}  & $0.7$ & $1.5$ & $0.8$  & $0.6$\\
 BKS \cite{4} & $0.9\pm0.08\pm0.21$ & $1.4\pm0.5\pm0.5$ &
$0.8\pm0.1\pm0.3$ & $0.6\pm0.1\pm0.2$\\
 LMMS \cite{5} & $0.63\pm0.08$ & $0.9\pm0.1$ & $0.53\pm0.03$ &
$0.2\pm0.2$\\
 BBD \cite{6} & $0.6$ & $1.1$ & $0.5$ & $0.6$\\
\hline
\hline
\end{tabular}

\vskip 7mm
\noindent
 Table 2. Calculated and measured $D\to K^*l\nu_l$ form factor ratios
$R_2=A_2(0)/A_1(0)$ and $R_V=V(0)/A_1(0)$.

\medskip
\begin{tabular}{lcc}
\hline
\hline
Ref.& $R_2(0)$ & $R_V(0)$\\
\hline
 Experiment &\hfill &\hfill\\
 E691 \cite{30} & $0.0\pm0.5\pm0.2$ & $2.0\pm0.6\pm0.3$\\
 E653 \cite{31} & $0.82^{+0.22}_{-0.23}\pm0.11$ & $2.00^{+0.34}_{-
0.32}\pm0.16$\\
 E687 \cite{32} & $0.78\pm0.18\pm0.10$ & $1.74\pm0.27\pm0.28$\\
\hline
 Theory &\hfill &\hfill\\
 RQM & $0.68$ & $0.98$\\
 ISGW \cite{1} & $1.0$ & $1.37$\\
 BSW \cite{2} & $1.36$ & $1.48$\\
 AW \cite{3} & $0.75$ & $1.87$\\
 BBD \cite{6} & $1.2$ & $2.2$\\
\hline
\hline
\end{tabular}

\vskip 7mm
\noindent
Table 3. The ratios $\Gamma(D\to K^*l\nu_l)/\Gamma(D\to Kl\nu_l)$ and
$\Gamma_L/\Gamma_T$ in comparison with the experimental data.

\medskip
\begin{tabular}{lcc}
\hline
\hline
Ref.& $\Gamma(K^*)/\Gamma(K)$ & $\Gamma_L/\Gamma_T$\\
\hline
 RQM & $0.65$ & $1.05$\\
 Experiment & &\\
 E691 \cite{30} & & $1.8^{+0.6}_{-0.4}\pm0.3$\\
 E653 \cite{31} & & $1.18\pm0.18\pm0.08$\\
 E687 \cite{32} & & $1.20\pm0.13\pm0.13$\\
 CLEO \cite{33} & $0.60\pm0.09\pm0.07$ & \\
 CLEO \cite{33} & $0.65\pm0.09\pm0.10$ & \\
\hline
\hline
\end{tabular}
\newpage
\noindent
Table 4. Measured and calculated ratios of form factors in
$D_s\to\phi l\nu_l$.

\medskip
\begin{tabular}{lcc}
\hline
\hline
Ref.& $R_2(0)$ & $R_V(0)$\\
\hline
 Exp. Average \cite{22} & $1.8\pm0.5$ & $2.0\pm0.7$\\
 CLEO \cite{A1} & $1.4\pm0.5\pm0.3$ & $0.9\pm0.6\pm0.3$\\
 E653 \cite{A2} & $2.1^{+0.6}_{-0.5}\pm0.2$ & $2.3^{+1.1}_{-
0.9}\pm0.4$\\
\hline
 Theory & &\\
 RQM & $0.55$ & $0.94$\\
 BKS \cite{4} & $2.0\pm0.19\pm0.23$ & $0.78\pm0.08\pm0.15$\\
 LMMS \cite{5} & $1.65\pm0.2$ & $0.33\pm0.36$\\
\hline
\hline
\end{tabular}
\vskip 7mm
\noindent
Table 5. Predicted and measured ratios of form factors in $B\to
D^*l\nu_l$
at $q^2=q^2_{max}$.

\medskip
\begin{tabular}{lcc}
\hline
\hline
Ref.& $R_2(q^2_{max})$ & $R_V(q^2_{max})$\\
\hline
 Experiment & &\\
 CLEO \cite{23} a) & $1.02\pm0.24$ & $1.07\pm0.57$\\
 CLEO \cite{23} b) & $0.79\pm0.28$ & $1.32\pm0.62$\\
\hline
 Theory & &\\
 RQM & $1.16$ & $1.74$\\
 ISGW \cite{1} & $1.14$ & $1.27$\\
 WSB \cite{2} & $1.06$ & $1.14$\\
 HQET-based \cite{24} & $1.26$ & $1.26$\\
 HQET-based \cite{25} & $1.14$ & $1.74$\\
\hline
\hline
\end{tabular}
\vskip 2cm
\rm
\noindent {\Large \bf Figure Captions}
\vskip 5mm
\noindent Fig.1 Lowest order vertex function.
\vskip 5mm
\noindent Fig.2 Vertex function with account of the quark
interaction. Dashed line
corresponds to the effective potential~(5). Bold line denotes the
negative-energy
part of quark propagator.
\vskip 5mm
\noindent Fig.3 Allowed region for semileptonic $B\to A(A^*)l\nu_l$
decay in terms of
the variables $w$ and $x$. Lower bound curve $w_m(x)$ is determined
by (\ref{74}), upper bound is $w_0=(M_A^2+M_B^2)/(2M_AM_B).$
\vskip 5mm
\noindent Fig.4 $(1/\Gamma)(d\Gamma/dx)$ for $B\to Dl\nu_l$ and $B\to
D^*l\nu_l$.
Absolute rates $d\Gamma/dx$ can be obtained by using
$\Gamma(D)=1.71\times10^{10}s^{-1}$
and $\Gamma(D^*)=2.92\times10^{10}s^{-1}$ for
$\tau_{B^0}=1.5\times10^{-12}$
and $|V_{bc}|=\,0.036$.
\vskip 5mm
\noindent Fig.5 $(1/\Gamma)(d\Gamma/dx)$ for $D\to Kl\nu_l$ and $D\to
K^*l\nu_l$.
Absolute rates $d\Gamma/dx$ can be obtained by using
$\Gamma(K)=6.68\times10^{10}s^{-1}$
and $\Gamma(K^*)=4.34\times10^{10}s^{-1}$ for
$\tau_{D^0}=0.415\times10^{-12}s$
, $\tau_{D^+}=1.06\times10^{-12}s$.
\vskip 5mm
\noindent Fig.6  $(1/\Gamma)(d\Gamma/dx)$ for $D_s\to\phi l\nu_l$.
Absolute rate $d\Gamma/dx$ can be obtained by using
$\Gamma=5.42\times10^{10}s^{-1}$
for $\tau_{D_s}=0.47\times10^{-12}s$.
\vskip 5mm
\newpage
\unitlength=1mm
\begin{picture}(150,150)
\large
\put(10,100){\line(1,0){50}}
\put(10,120){\line(1,0){50}}
\put(35,120){\circle*{5}}
\multiput(32.5,130)(0,-10){2}{\begin{picture}(5,10)
\put(2.5,10){\oval(5,5)[r]}
\put(2.5,5){\oval(5,5)[l]}\end{picture}}
\put(5,120){$b$}
\put(5,100){$\bar q$}
\put(5,110){$B$}
\put(65,120){$a$}
\put(65,100){$\bar q$}
\put(65,110){$A$}
\put(43,140){$W$}
\put(30,85){\Large\bf Fig. 1}
\put(10,20){\line(1,0){50}}
\put(10,40){\line(1,0){50}}
\put(25,40){\circle*{5}}
\put(25,40){\thicklines \line(1,0){20}}
\multiput(25,40.5)(0,-0.1){10}{\thicklines \line(1,0){20}}
\put(25,39.5){\thicklines \line(1,0){20}}
\put(45,40){\circle*{1}}
\put(45,20){\circle*{1}}
\multiput(45,40)(0,-4){5}{\line(0,-1){2}}
\multiput(22.5,50)(0,-10){2}{\begin{picture}(5,10)
\put(2.5,10){\oval(5,5)[r]}
\put(2.5,5){\oval(5,5)[l]}\end{picture}}
\put(5,40){$b$}
\put(5,20){$\bar q$}
\put(5,30){$B$}
\put(65,40){$a$}
\put(65,20){$\bar q$}
\put(65,30){$A$}
\put(33,60){$W$}
\put(90,20){\line(1,0){50}}
\put(90,40){\line(1,0){50}}
\put(125,40){\circle*{5}}
\put(105,40){\thicklines \line(1,0){20}}
\multiput(105,40.5)(0,-0.1){10}{\thicklines \line(1,0){20}}
\put(105,39,5){\thicklines \line(1,0){20}}
\put(105,40){\circle*{1}}
\put(105,20){\circle*{1}}
\multiput(105,40)(0,-4){5}{\line(0,-1){2}}
\multiput(122.5,50)(0,-10){2}{\begin{picture}(5,10)
\put(2.5,10){\oval(5,5)[r]}
\put(2.5,5){\oval(5,5)[l]}\end{picture}}
\put(85,40){$b$}
\put(85,20){$\bar q$}
\put(85,30){$B$}
\put(145,40){$a$}
\put(145,20){$\bar q$}
\put(145,30){$A$}
\put(133,60){$W$}
\put(70,5){\Large \bf Fig. 2}

\end{picture}

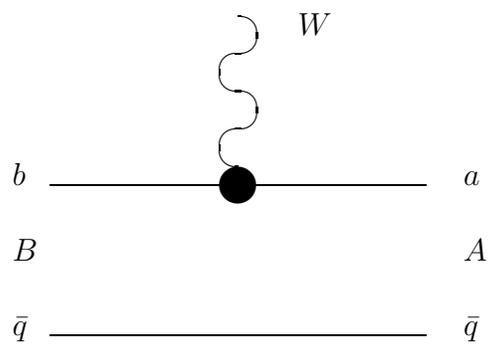
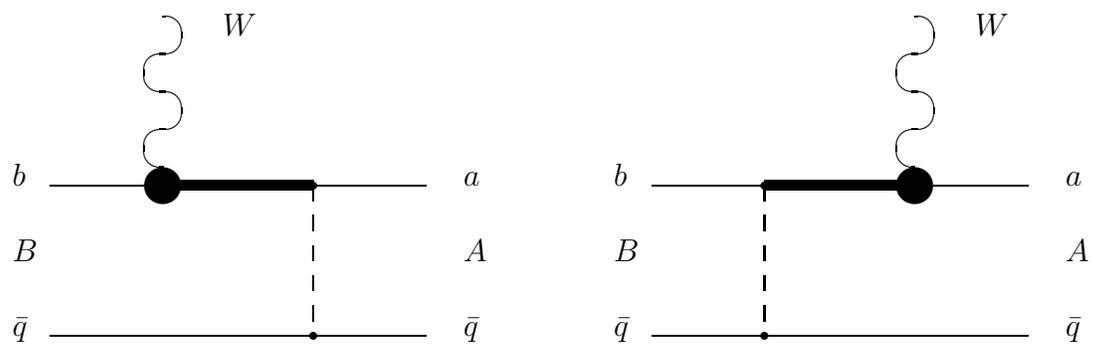
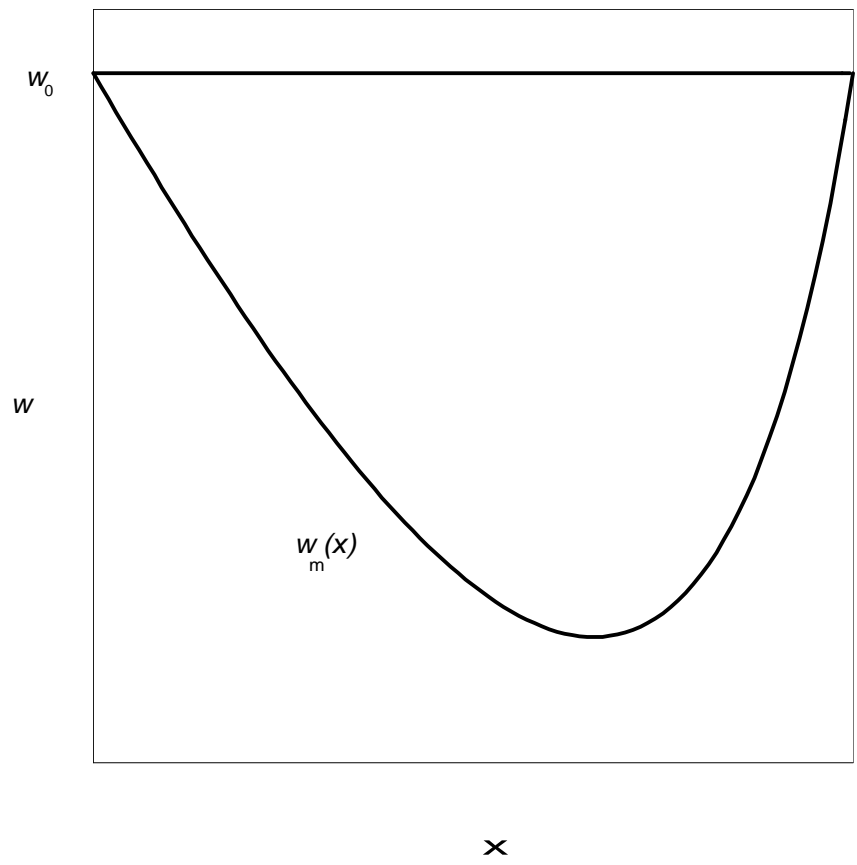
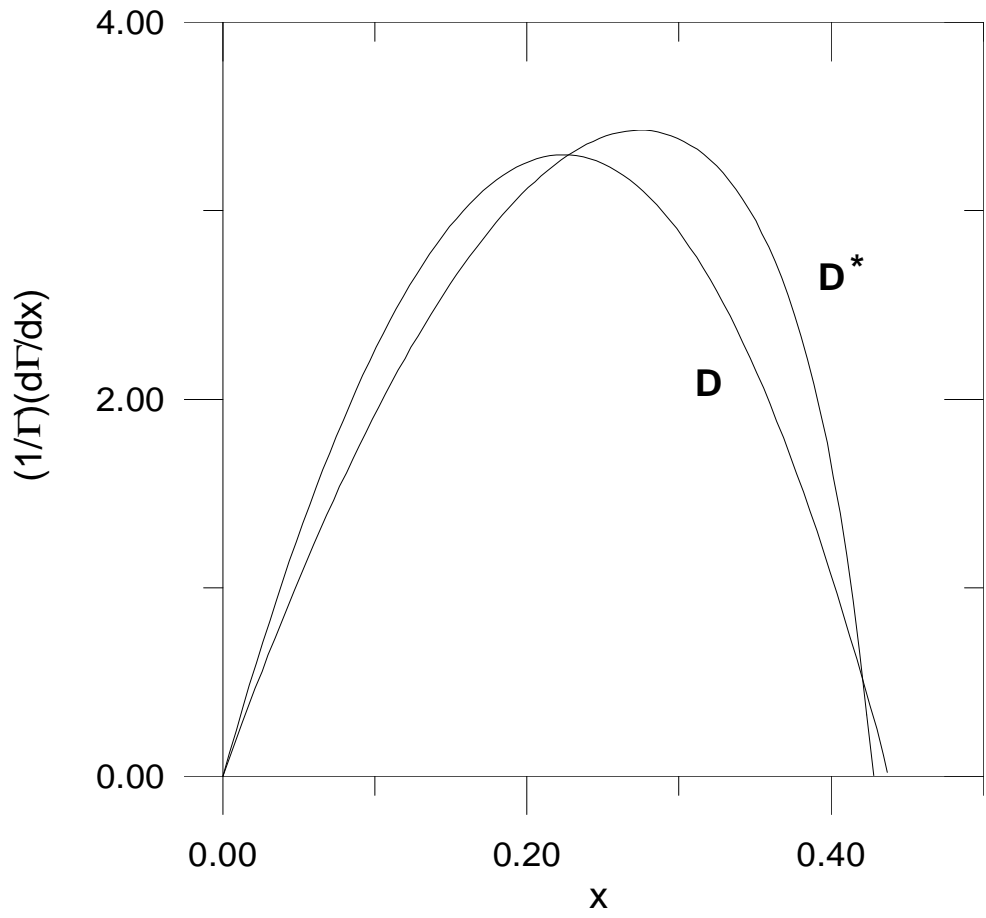
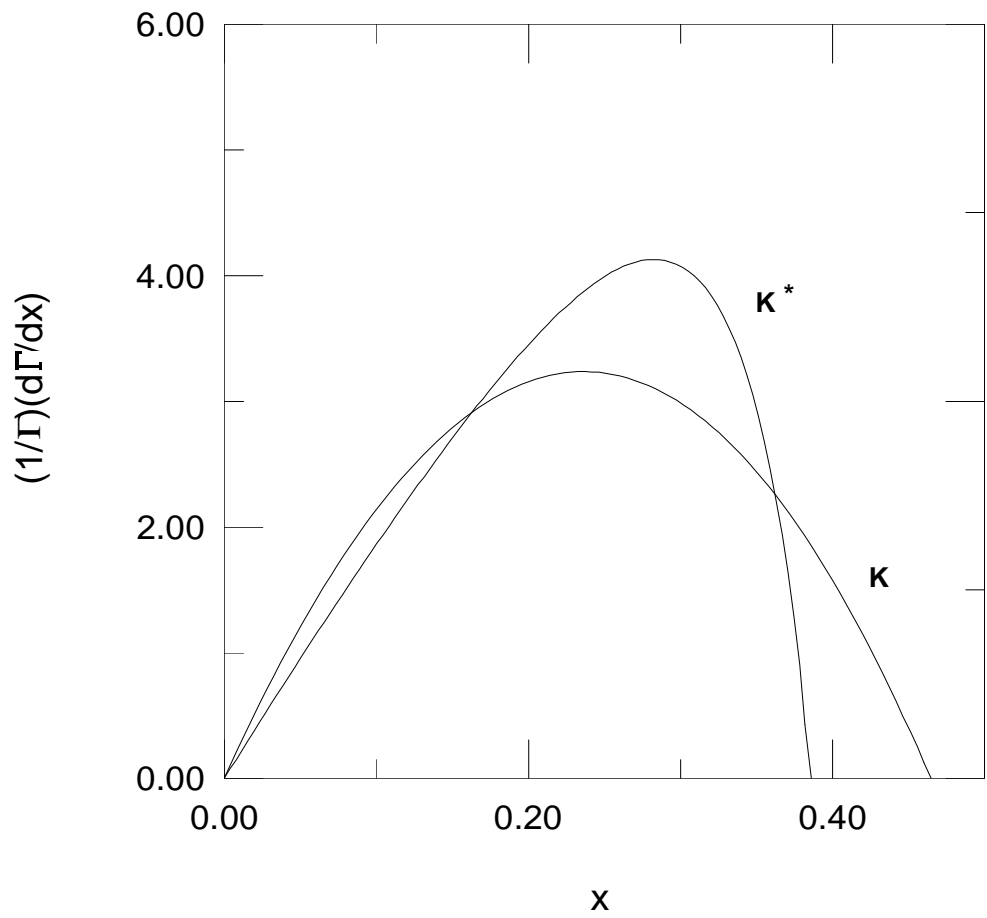
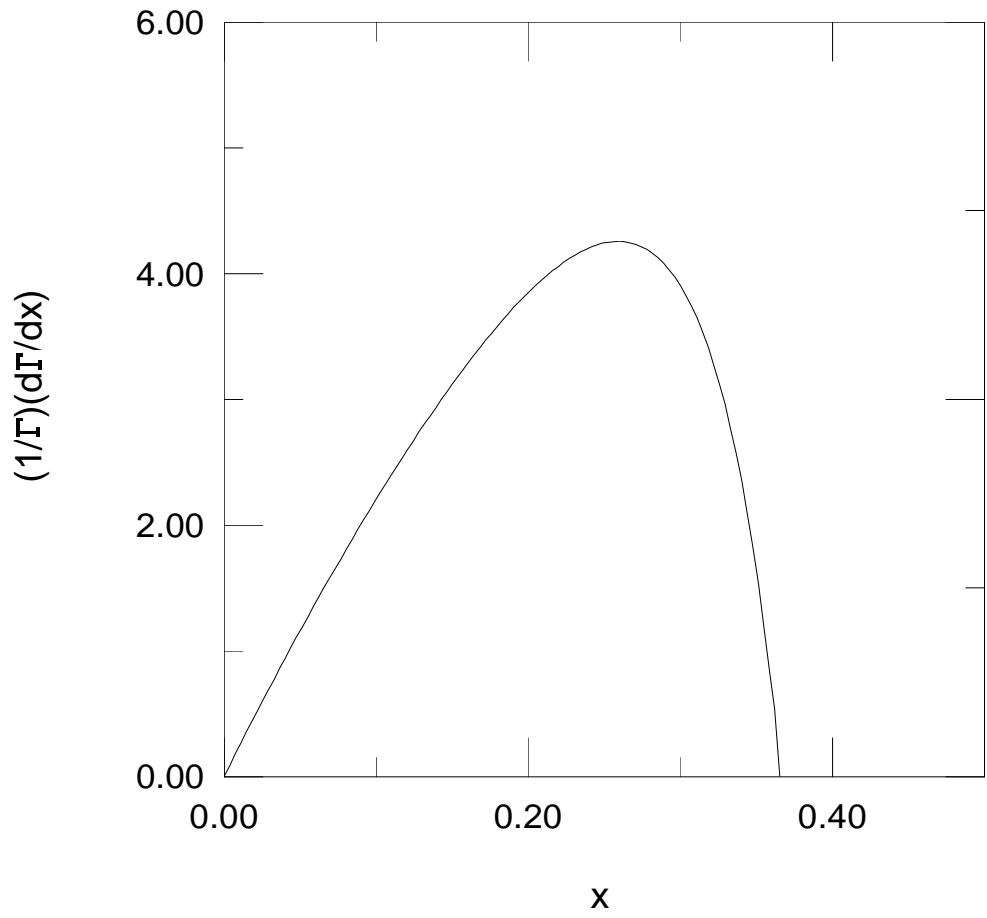
\end{document}